\documentclass[preprint2]{aastex}

\def \sun {\hbox {$\odot$}}

\def \ref {\reference{}}

\def \hcm {\hbox {\ifmmode $ atoms cm$^{-2}\else atoms cm$^{-2}$\fi}}

\def\approxgt{\mathrel{\hbox{\rlap{\lower.55ex \hbox {$\sim$}}
  \kern-.3em \raise.4ex \hbox{$>$}}}}
\def\approxlt{\mathrel{\hbox{\rlap{\lower.55ex \hbox {$\sim$}}
  \kern-.3em \raise.4ex \hbox{$<$}}}}
\shorttitle{Z-diagram of LMC~X--2}
\shortauthors{Smale et al.}

\received{Accepted for APJ Main Journal}
\begin{document}

\title{The complete $Z$-diagram of LMC~X--2}

\author{A. P. Smale\altaffilmark{1}, 
J. Homan\altaffilmark{2},
and E. Kuulkers\altaffilmark{3}
}

\altaffiltext{1}{USRA Research Scientist, Laboratory for High Energy
Astrophysics, Code 660.1, NASA/Goddard Space Flight Center, Greenbelt,
MD 20771; alan@osiris.gsfc.nasa.gov}

\altaffiltext{2}{Osservatorio Astronomico di Brera, Via
E. Bianchi 46, 23807 Merate LC, Italy; homan@merate.mi.astro.it}

\altaffiltext{3}{ESA-ESTEC, Science Operations \& Data Systems
Division, SCI-SDG, Keplerlaan 1, 2201 AZ Noordwijk, The Netherlands;
ekuulker@rssd.esa.int}

\begin{abstract}

We present results from four {\it Rossi X-ray Timing Explorer (RXTE)}
observations of the bright low mass X-ray binary LMC~X--2. During
these observations, which span a year and include over 160~hrs of
data, the source exhibits clear evolution through three branches on
its hardness-intensity and color-color diagrams, consistent with the
flaring, normal, and horizontal branches (FB, NB, HB) of a $Z$-source,
and remarkably similar to $Z$-tracks derived for GX~17$+$2, Sco X--1
and GX~349$+$2.  LMC~X--2 was observed in the FB, NB, and HB for
roughly 30\%, 40\%, and 30\% respectively of the total time covered.
The source traces out the full extent of the $Z$ in $\sim$1 day, and
the $Z$-track shows evidence for secular shifts on a timescale in
excess of a few days. Although the count rate of LMC~X--2 is low
compared with the other known $Z$-sources due to its greater distance,
the power density spectra
selected by branch show very-low-frequency noise characteristics at
least consistent with those from other $Z$-sources. We thus confirm
the identification of LMC~X--2 as a $Z$-source, the first identified
outside our Galaxy.

\end{abstract}


\section{Introduction}

Accreting low mass X-ray binaries (LMXBs) have been divided into two
groups based on their broadband spectral and timing properties. The
six currently-known $Z$-sources, named for the three-branched shape
they describe on a diagram of 'hard' color versus 'soft' color, are
Sco~X--1, GX~17$+$2, GX~349$+$2 (=Sco~X--2), Cyg~X--2, GX~340$+$0, and
GX~5$-$1 (Hasinger \& van der Klis 1989). From the top down, the three
branches of the $Z$ are known as the horizontal branch (HB), normal
branch (NB), and flaring branch (FB), and the power spectra of these
sources generally show several types of quasi-periodic oscillations
(QPOs) and noise components (e.g. van der Klis 1995) based on the
position of the source on the $Z$. When bright, Cir~X--1 also shows
$Z$-source characteristics (Shirey et al.\ 1998; Shirey, Bradt \&
Levine 1999). The mass accretion rate is generally thought to increase as the
source passes from the HB, through the NB, to the FB. Motion along
the $Z$ is continuous, with no jumps from branch to branch.

The remainder of the bright Galactic LMXBs are known as 'atoll' sources and
describe a simpler curved shape on a color-color diagram, generally
consisting of a band of points, plus one or more 'islands'. Atoll
sources that display a range of intensities approaching two orders of
magnitude may reveal a three-band shape (Muno, Remillard
\& Chakrabarty 2002), but on much longer timescales
($\sim$weeks--months) than the $Z$-source tracks
($\sim$hours--days). The difference
in their nature implies that $Z$- and atoll-sources differ in their
evolutionary history (Hasinger \& van der Klis 1989).

A previous {\it RXTE} observation of LMC~X--2 in 1997 December showed
timing and spectral characteristics more typical of a $Z$-source in
its flaring branch than of an atoll source in any intensity state
(Smale \& Kuulkers 2000). Based on these findings and the high
intrinsic luminosity of LMC~X--2, which historically spans
0.4--2.0$L_{Edd}$ for reasonable estimates of the neutron star mass,
we then provisionally reclassified LMC~X--2 as a $Z$-source, the
seventh known and the first to be detected beyond our Galaxy. Here
we present extensive additional {\it RXTE} data from LMC~X--2
in which the source convincingly traces out all three branches in its
hardness-intensity and 
color-color diagrams, confirming its $Z$-source identification.

\section{Observations and Analysis}

LMC~X--2 was observed using the {\it Rossi X-ray Timing Explorer
(RXTE)} (Bradt, Rothschild, \& Swank 1993) on four occasions during
2001--2002 (Table~1), for a total on-source good time of 600~ksec.
The data presented here were collected by the proportional counter
array (PCA) instrument using the Standard 2 and Good Xenon
configurations, with time resolutions of 16~s and 0.95~$\mu$s
respectively.  The PCA consists of five Xe proportional counter units
(PCUs) numbered 0--4, with a combined total effective area of about
6500~cm$^2$ (Jahoda et al.\ 1996). PCUs 0 and 2 were reliably on
during all observations, and for consistency across the whole dataset
we utilized data from just these PCUs for the generation of light
curves and color-color diagrams. To maximize the statistics, for our
timing analysis we used data from all available PCUs.

\begin{deluxetable}{cccc}     %% Table 1
\tablewidth{0pc}
\tablecolumns{4}
\tablecaption{
OBSERVATIONS OF LMC~X--2}
\tablehead{
\colhead{Obs.}&
\colhead{ObsID}&
\colhead{Date/Time (UT)}&
\colhead{Good Time}
}
\startdata
1 & 50041-01-01-* & 2001 Feb 10/17:35 -- Feb 14/03:28 & 149 ksec \\
2 & 60017-01-01-* & 2001 Aug 30/12:12 -- Sep 03/02:40 & 143 ksec \\
3 & 60017-01-02-* & 2001 Dec 13/22:22 -- Dec 16/23:27 & 139 ksec \\
4 & 60017-01-03-* & 2002 Feb 01/02:56 -- Feb 06/16:21 & 149 ksec \\
\enddata
\end{deluxetable}

The background-subtracted light curves for each observation presented
in Figure 1 include LMC~X--2 data from the energy range 2.5--18.5~keV.
To construct the hardness-intensity diagrams (HID) and color-color
(CD) diagrams also shown, we followed conventional practice in the
field and defined several ``colors'', representing the ratios of count
rates in two different energy bands. The ``overall'' color is the
ratio of the 6.5--18.5~keV band to the 2.5--6.5~keV band, and the HID
are plots of this overall color against the 2.5--18.5~keV source
intensity.  CD are plots of the ``hard'' color
[(9.8--18.5~keV)/(6.5--9.8~keV] against the ``soft'' color
[(4.5--6.5~keV)/(2.5--4.5~keV)].

\section{Results}

\subsection{Light curves and colors}

In Figure~1a we show the background-subtracted light curve for the
2001 February observation. Two episodes of major flaring are observed,
during which the source intensity almost doubles. Preceding this
flaring behavior are two sections of low-level variability, and
following the second series of flares the source settles into an
interval of slightly higher, almost constant count rate.

Figure~1a also shows the HID and CD derived for these data using the
colors defined in the previous section. These diagrams bear a
remarkable resemblance to those seen in e.g. GX~17$+$2
(Hasinger \& van der Klis 1989, Kuulkers et al. 1997, Homan et
al. 2002) and show very little similarity with those seen from atoll
sources (e.g.\ Hasinger \& van der Klis 1989, Prins \& van der Klis
1997, Wijnands et al. 1998). Dividing the data into sections of a
few hundred seconds each and
examining the correlated variability in each section we can 
identify the three characteristic branches of the $Z$;
during the 83-hr span of the observation, LMC~X--2 moves in turn
through the NB, FB, NB, FB, NB, and HB.

Subsequent panels of Figure~1 show the light curves, HIDs, and CDs for
the 2001 August, 2001 December, and 2002 February observations of
LMC~X--2. The source spends almost all of the second observation on
the FB, with some data defining the NB/FB vertex.  The third
observation contains data from the NB and part of the HB, while the
fourth observation provides evidence for strong evolution along the
HB, with a swift ($\sim$few thousand second) excursion through the NB
and a brief flare lasting $\sim$10,000s.  Examination of the
hardness-intensity and color-color information for each observation on
a point by point basis shows that the source moves along each branch
in a continuous progression.  No bursts were detected during any of
the observations.

In Figure~2 we present the HID and CD for all four LMC~X--2
observations combined.  The combined CD (Fig.~2b) is a simple
superposition of the data from all four of the individual CDs in
Figure 1. However, a small secular shift in the position of the HB/NB
and NB/FB vertices is evident in the individual HIDs, and we thus
subtracted constants of 0.045 and 0.01 from the colors
given in Fig.~1b and Fig.~1d respectively, prior to the
superposition. (Failure to do this results in a combined HID with an
unrealistically broad and skewed Normal branch, and the appearance of
two FBs.)

To assess whether these shifts were intrinsic to the source or to
the {\it RXTE} PCA detectors, we analyzed data from the Crab obtained close
in time to our LMC~X--2 observations. Extraction of the Crab CD and
HID, using the same channels as those used for LMC~X--2, showed that
the gradual gain change of the PCA detectors over the span of our
observations could account for a variation of less than 1\% in the derived
colors. The secular shifts in the HIDs of LMC~X--2 amount to 10\% and 2.5\%
changes, greatly in excess of this. We thus conclude that the source
itself displayed intrinsic shifts, similar to the secular shifts seen
in other $Z$-sources (see Discussion).

\subsection{Variability analysis}

For our power spectral analysis data were divided into three
selections representing the HB, NB, and FB, based on their position
along the Z-track in the CD and HID. Power spectra were created from
the high-time-resolution Good Xenon mode data covering 2--21 keV, 
using standard Fast
Fourier Transform techniques over the frequency range 1/128--512~Hz.
No narrow features were present in the power density spectra, and each
could be well fit ($\chi^2_\nu$=1.06--1.21) 
with a single power law with shape P $\propto$ $\nu^{-\gamma}$, 
representing very low frequency noise (VLFN). The low
count rate of the source does not allow us to parameterize
high-frequency noise (HFN) or perform a sensitive search for QPOs.

The horizontal branch spectrum (Figure~3a) was constructed using
170~ksec of data, and the VLFN component has a power law index of
$\gamma$=0.60$\pm$0.05 and an rms amplitude of 1.9$\pm$0.1\%
(0.01--1~Hz). For the normal branch spectrum (Fig.~3b) 229~ksec of
data were used, and the power law component has
$\gamma$=0.90$\pm$0.05, rms=1.8$\pm$0.1\%.  The flaring branch
spectrum uses 161~ksec of data, and has $\gamma$=1.33$\pm$0.03,
rms=3.20$\pm$0.05\%. Sections of data totaling 40~ksec were excluded
from the analysis due to ambiguity about which branch they fell into;
datapoints obtained around the NB/FB vertex are particularly difficult
to assign.

For the data obtained on the NB we detected no QPOs, and and derive
upper limits of 1\% rms for oscillations with parameters typical of
$Z$-sources (e.g.\ centroid position of $\sim$6~Hz, FWHM 2~Hz). For
the FB data we detected a peak at 13.2$\pm$0.2~Hz with width
4.9$\pm$0.3~Hz and rms 1.9$\pm$0.5\%, but its low (2$\sigma$)
significance was not compelling.

\section{Discussion}

Using 600~ksec of new data from the bright LMC binary LMC~X--2, we
have shown that the source traces out $Z$-like tracks in the HID and
CD, moving continuously along three branches and not jumping
instantaneously from one to the next. The pattern, width, and extent
of the tracks show particular similarity with those observed from
GX~17$+$2 (e.g.\ Homan et al.\ 2002), Sco~X--1 (e.g.\ Dieters \& van
der Klis 2000), and GX~349$+$2 (e.g.\ O'Neill et al.\ 2002), and have
little in common with the simpler atoll source tracks.

By convention, position on the $Z$ is parameterized using the $S_z$
rank number, in which the HB/NB and NB/FB vertices are defined as
$S_z$=1 and $S_z$=2 respectively, and the rest of the track is scaled
according to the length of the NB (e.g.\ Hasinger et al.\ 1990; Hertz
et al.\ 1992; Dieters \& van der Klis 2000). Adopting this
parameterization, we have obtained data from LMC~X--2 covering the
range $S_z=-$1.0--4.6, a full span equivalent to that seen in the
other $Z$-sources.  Our observations show good coverage of all three
branches; LMC~X--2 spent roughly 30\%, 40\%, and 30\% of the total
observing time in the HB ($S_z \leq 1$), NB ($1 < S_z \leq 2$), and FB
($S_z > 2$) respectively.

The power density spectra derived from the three branches of LMC~X--2
appear rather featureless when compared to those of the $Z$ sources,
or indeed to those of the atoll sources. The spectra are dominated by
VLFN, with little significant evidence for either peaked low frequency
noise (LFN) at 10--100~Hz, or NBO/FBO.  The power law indices and rms
amplitudes measured for the VLFN in LMC~X--2, and the overall
steepening trend of the VLFN when moving from the HB to the FB, are
similar to those observed in the $Z$ sources GX~349$+$2, Sco~X--1, and
GX~17$+$2, with a particularly close match to the relatively flat VLFN
indices of $\gamma$=0.6--2.0 measured in GX~17$+$2 data (Kuulkers et
al.\ 1997; Dieters \& van der Klis 2000; O'Neill et al.\ 2001, 2002;
Homan et al.\ 2002).  Of course, the power law indices we measure may
be artificially low, due to the possibility of 'contamination' by
unresolved LFN and HFN components.

On the basis of their timing and spectral behavior, the Galactic
$Z$-sources have been divided into two categories (Hasinger \& van der
Klis 1989; Penninx et al.\ 1991; Kuulkers et al. 1994). It has been
proposed that the 'Cyg-like' sources, Cyg~X--2, GX~5--1, and
GX~340$+$0, have either a higher inclination (Kuulkers, van der Klis
\& Vaughan 1996) or a higher magnetic field strength (Psaltis, Lamb \&
Miller 1995) than the 'Sco-like' sources, Sco~X--1, GX~17$+$2, and
GX~349$+$2.  Based on the shapes of its HID and CD, and the
observation of intensity flares rather than dips in its FB, LMC~X--2
falls squarely into the Sco-like category.  However, the Cyg-like
sources typically show secular variations in the position of the $Z$
of $\sim$5--10\% (GX~5--1 -- Kuulkers et al.\ 1994; Cyg X--2 --
Kuulkers et al.\ 1996, Wijnands et al.\ 1997b; GX340$+$0 -- Kuulkers
\& van der Klis 1996), while in the Sco-like sources secular changes
in the $Z$ are rare, seen only occasionally in one source, GX~17$+$2
(Wijnands et al 1997a; Homan et al 2002). In this context, the secular
shifts of 2.5--10\% we observe from LMC~X--2 over the course of a year
become intriguing. A series of {\it RXTE} observations is currently
under way to provide medium-term monitoring of LMC~X--2 to further
study the degree and timescale of these secular variations, and
perform additional sensitive searches for QPOs.

\clearpage

\begin{figure*}[!ht]                   %Fig 1a
\begin{center}
\includegraphics[height=150mm,angle=270]{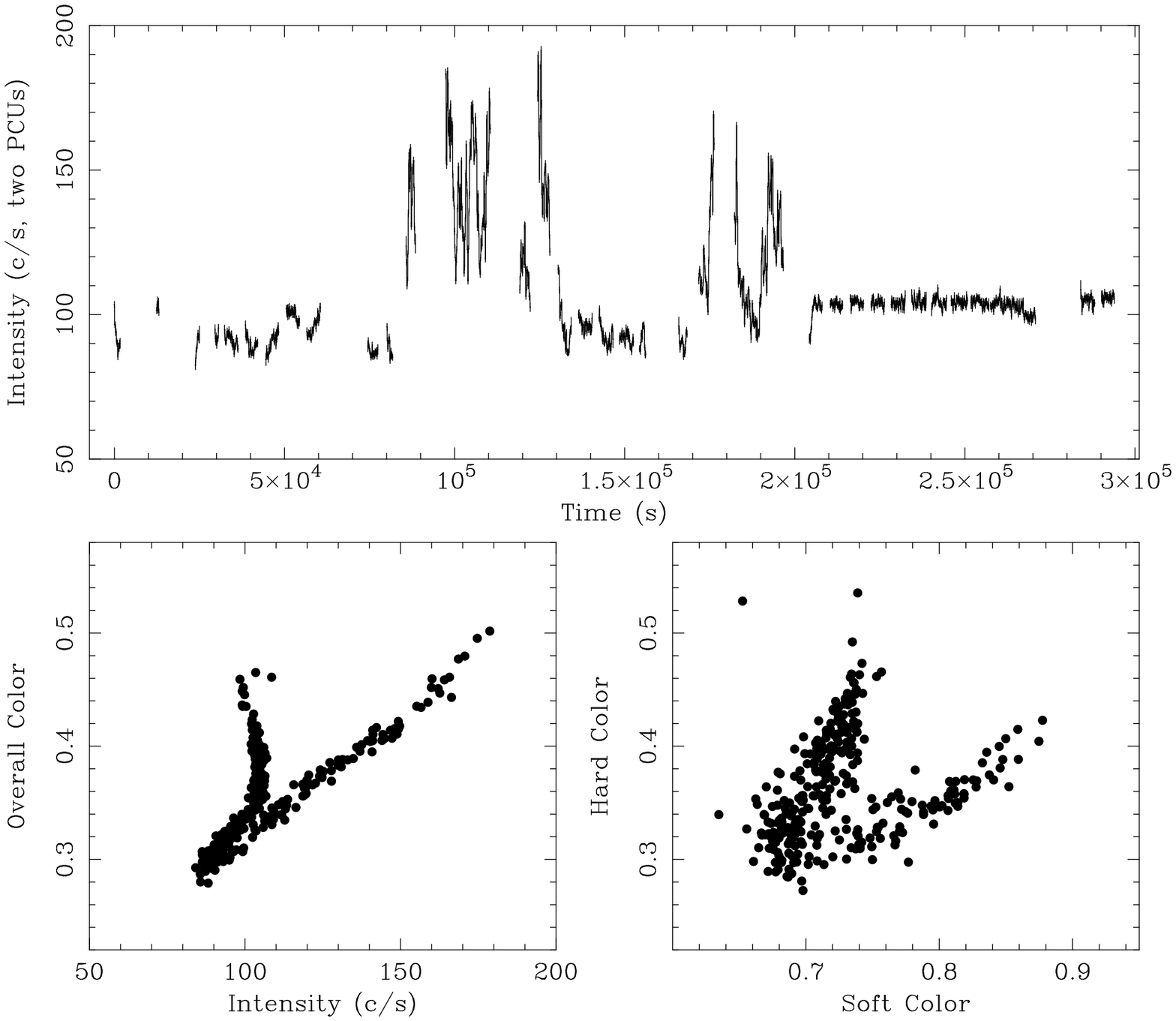} 
\figurenum{1}
\caption{
{\it RXTE} PCA light curves, HIDs, and CDs of LMC~X--2 obtained
during each of the four observations: 
(a) 2001 February 10--14;
(b) 2001 August 30--September 2;
(c) 2001 December 13--16;
(d) 2002 February 1--6.
The definitions of the energy ranges used for the
colors can be found in Section 2. The binsize is 128s for the light
curve, and 512s for the HID and CD. Data points
from the four observations are indicated using filled circles,
crosses, squares, and triangles respectively, for easy comparison with
Figure~2.
}
\end{center}
\end{figure*}

\begin{figure*}[!ht]                   %Fig 1b
\begin{center}
\includegraphics[height=150mm,angle=270]{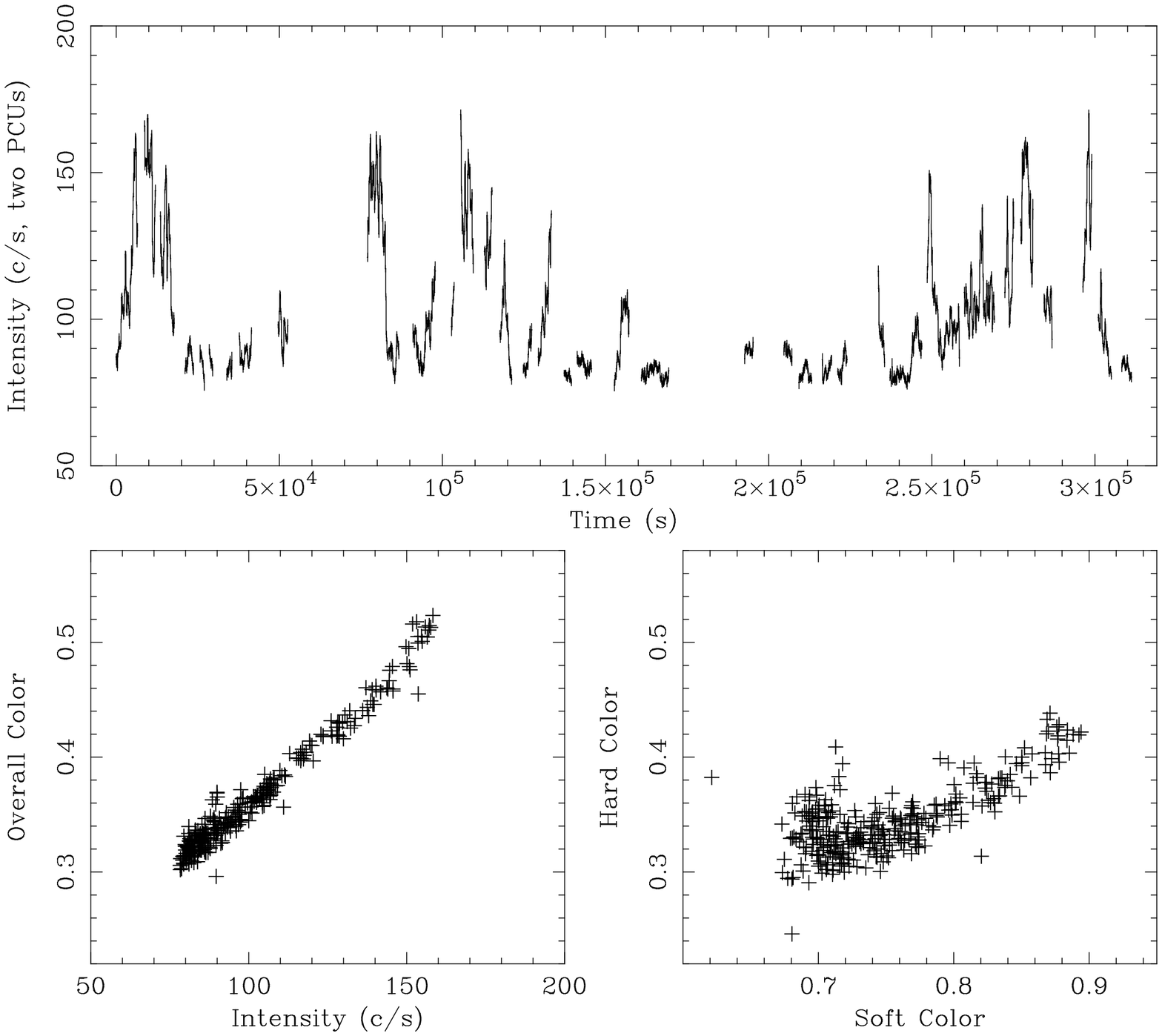} 
\figurenum{1b}
\caption{
}
\end{center}
\end{figure*}

\begin{figure*}[!ht]                   %Fig 1c
\begin{center}
\includegraphics[height=150mm,angle=270]{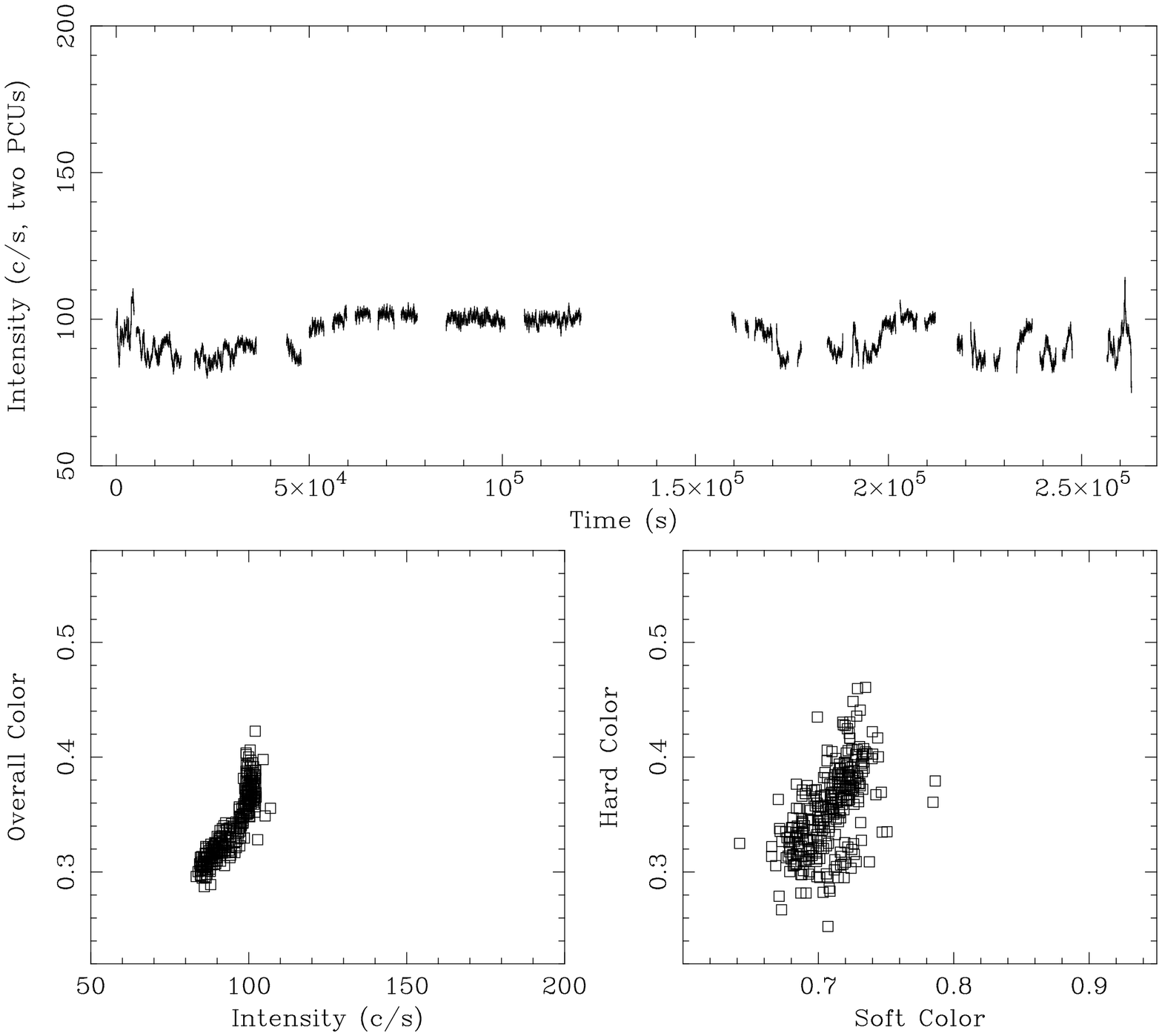} 
\figurenum{1c}
\caption{
}
\end{center}
\end{figure*}

\begin{figure*}[!ht]                   %Fig 1d
\begin{center}
\includegraphics[height=150mm,angle=270]{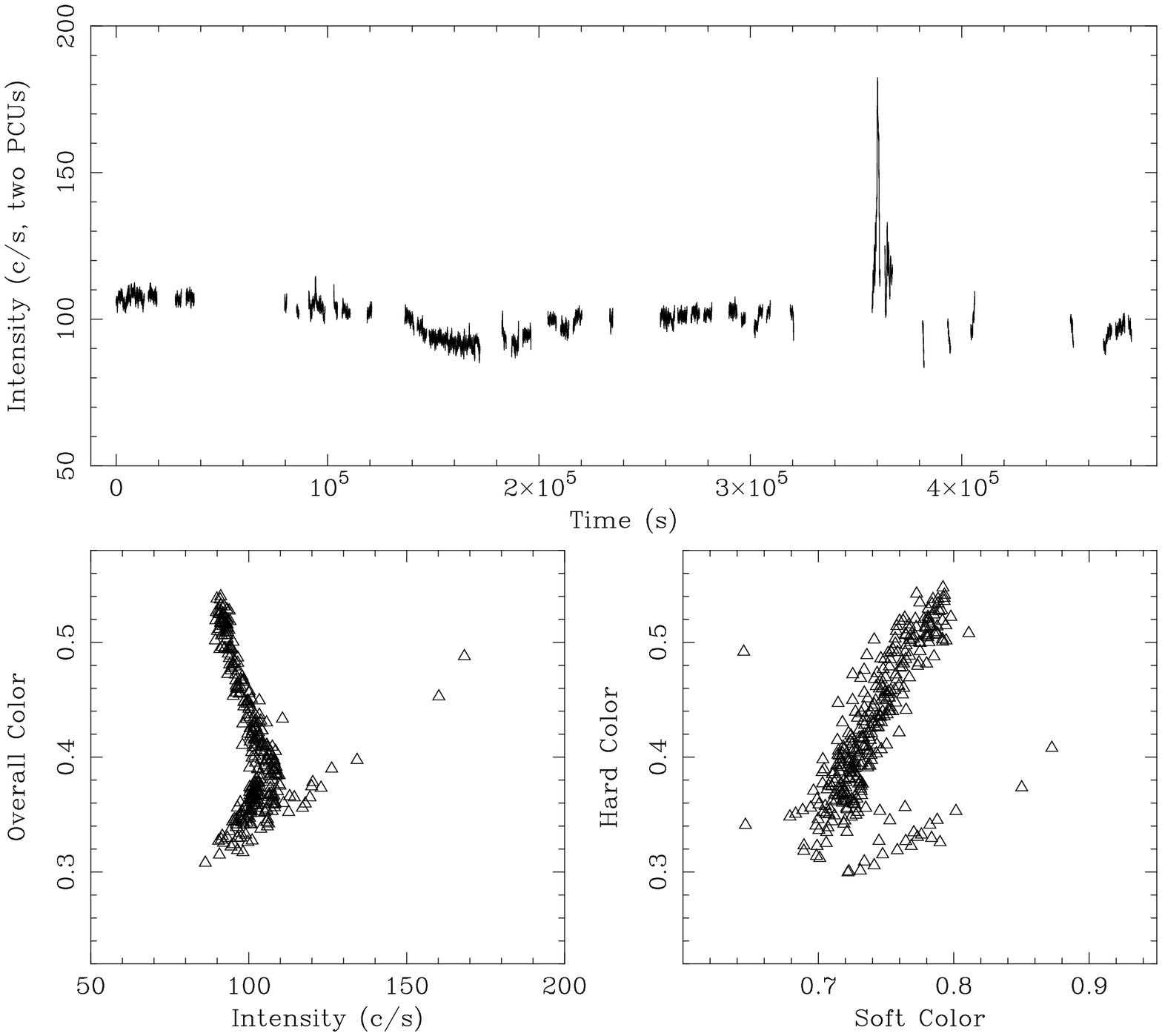} 
\figurenum{1d}
\caption{
}
\end{center}
\end{figure*}

\begin{figure*}[!ht]                   %Fig 2
\begin{center}
\includegraphics[width=60mm,height=78mm,angle=270]{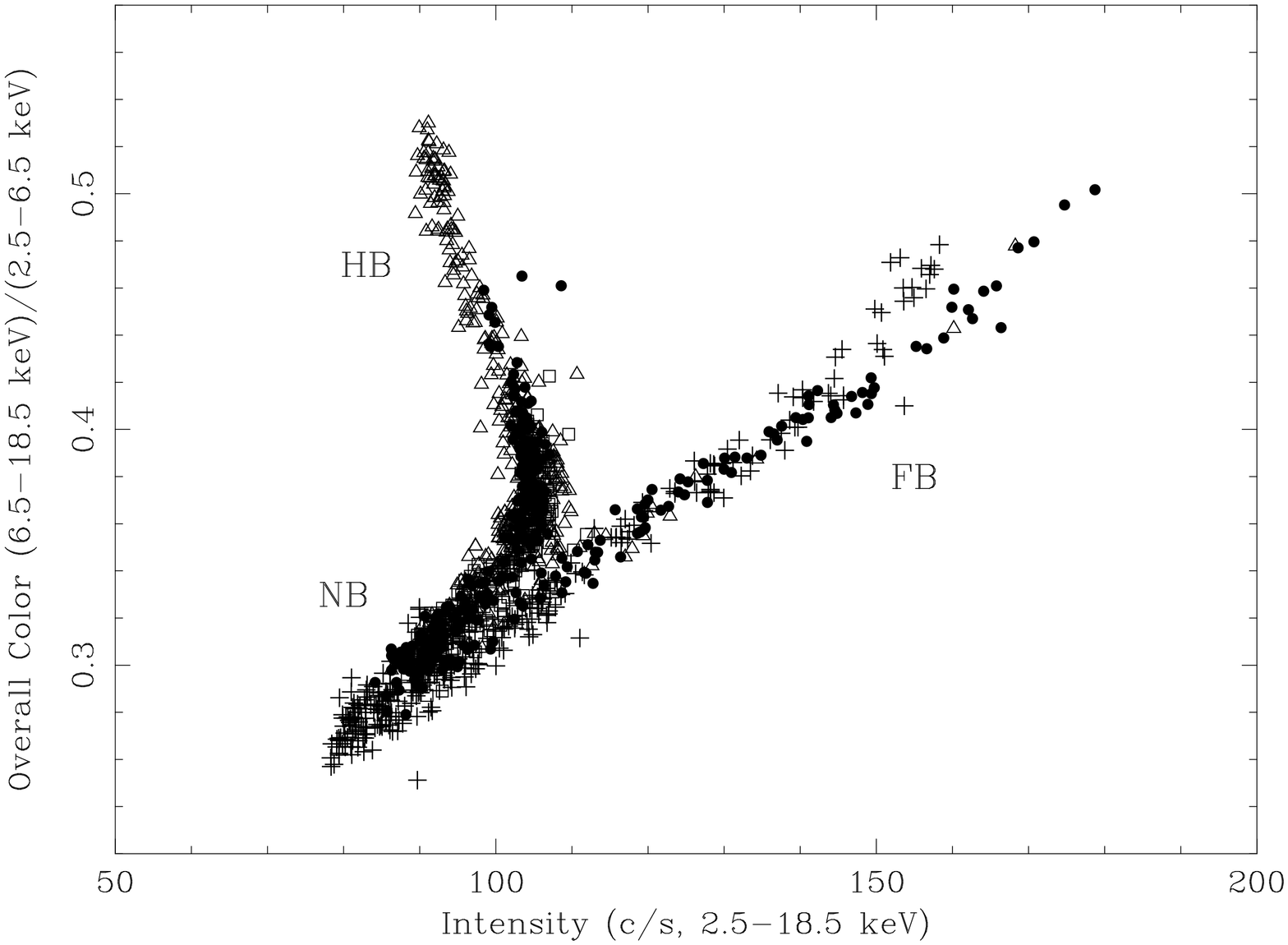} 
\includegraphics[width=60mm,height=78mm,angle=270]{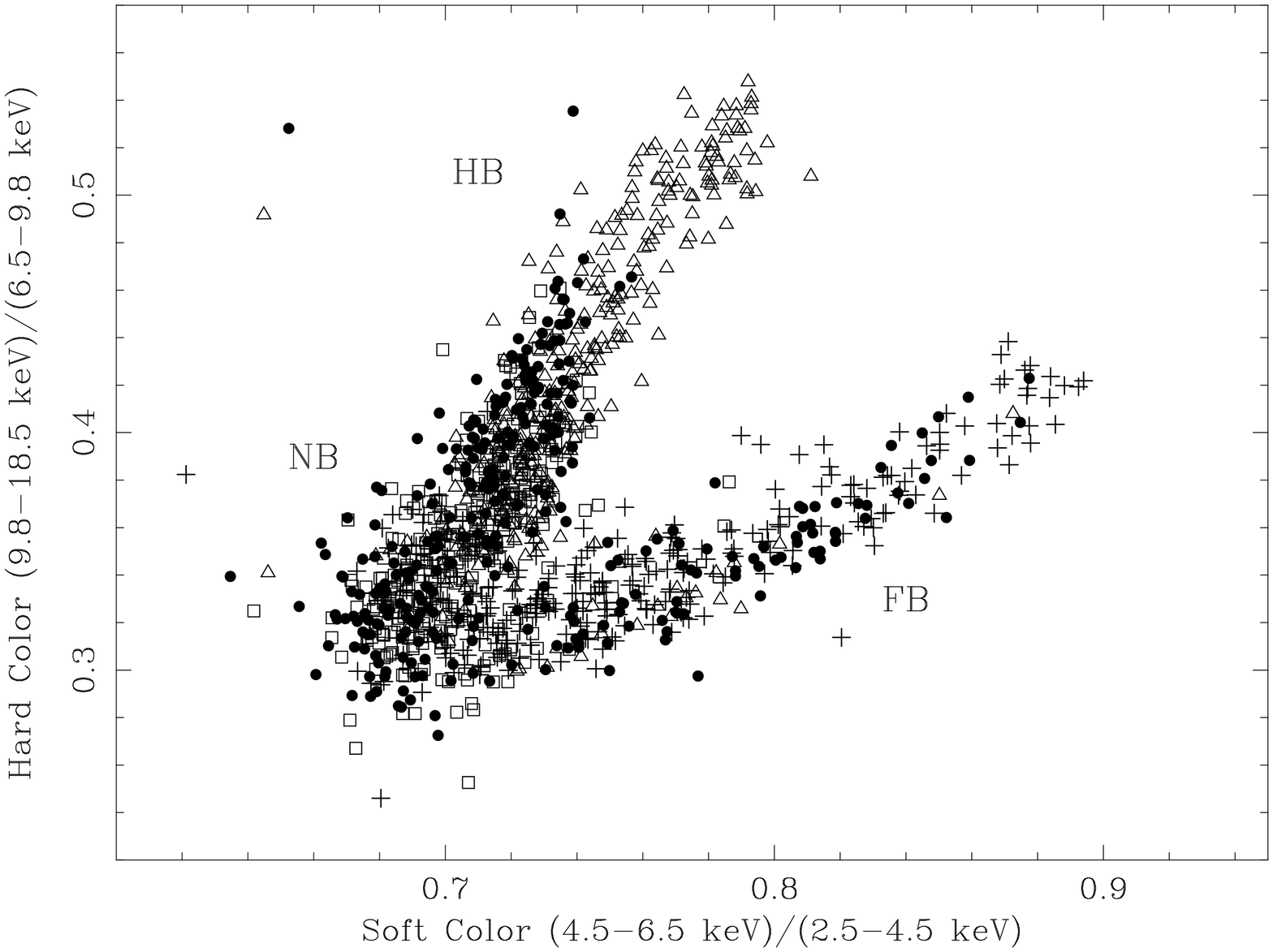} 
\figurenum{2}
\caption{
Left: Combined HID for all four observations of LMC~X--2. Right:
Combined CD. Each datapoint represents 512s of data.  Data points
from each of the four observations are indicated using filled circles,
crosses, squares, and triangles respectively, matching the convention
in Figure~1. Compare particularly with the similar figures for
GX~17$+$2: Homan et al.\ (2002), Fig.~1.
}
\end{center}
\end{figure*}

\begin{figure*}[!ht]                   %Fig 3
\begin{center}
\includegraphics[height=60mm,angle=0]{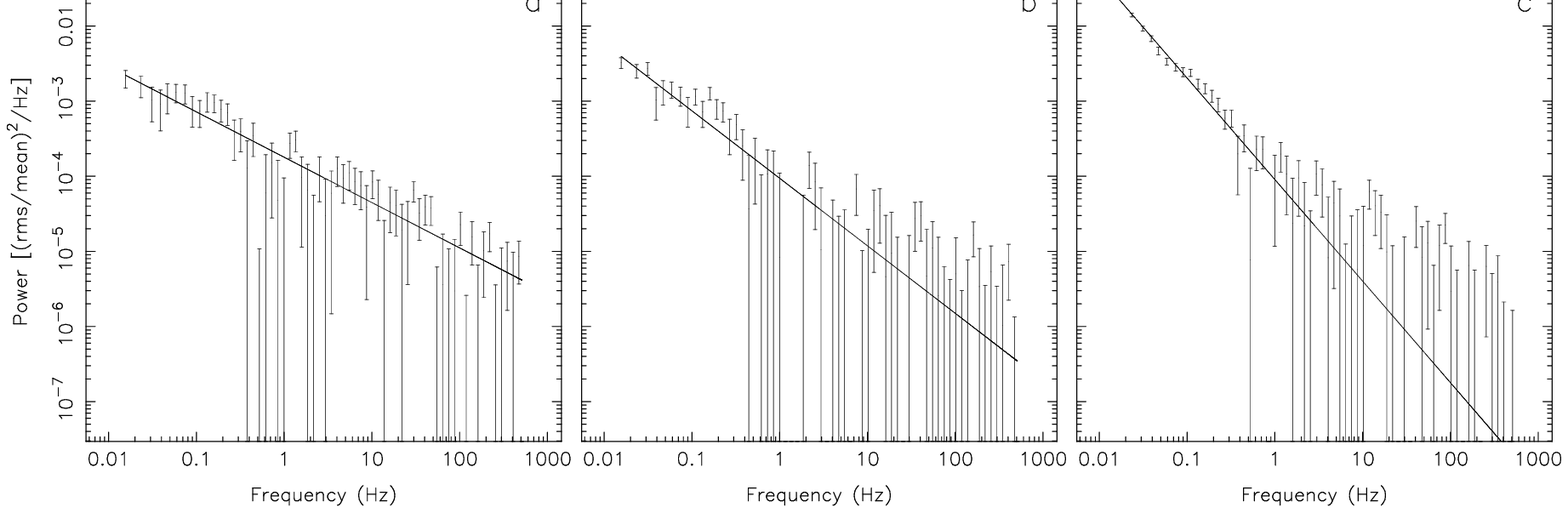}
\figurenum{3}
\caption{
Power density spectra of LMC~X--2 for the three branches of the Z: 
(a) Horizontal branch, (b) Normal branch, (c) Flaring branch.
}
\end{center}
\end{figure*}

\end{document}